\documentclass[letter]{aa}
\def\folio{\ifnum\pageno=1\nopagenumbers\else\number\pageno\fi}

\def\lax    {\ifmmode{_<\atop^{\sim}}\else{${_<\atop^{\sim}}$}\fi}
\def\gax    {\ifmmode{_>\atop^{\sim}}\else{${_>\atop^{\sim}}$}\fi}
\newbox\grsign      \setbox\grsign=\hbox{$>$} 
\newdimen\grdimen   \grdimen=\ht\grsign
\newbox\simgreatbox \setbox\simgreatbox=\hbox{\raise.5ex\hbox{$>$}\llap
                        {\lower.5ex\hbox{$\sim$}}}\ht1=\grdimen\dp1=0pt
\newbox\simlessbox  \setbox\simlessbox =\hbox{\raise.5ex\hbox{$<$}\llap
                        {\lower.5ex\hbox{$\sim$}}}\ht2=\grdimen\dp2=0pt

\def\d {\phantom{$0$}}
\def\dd {\phantom{$00$}}

\newbox\grsign \setbox\grsign=\hbox{$>$} \newdimen\grdimen \grdimen=\ht\grsign
\newbox\laxbox \newbox\gaxbox
\setbox\gaxbox=\hbox{\raise.5ex\hbox{$>$}\llap
     {\lower.5ex\hbox{$\sim$}}}\ht1=\grdimen\dp1=0pt
\setbox\laxbox=\hbox{\raise.5ex\hbox{$<$}\llap
     {\lower.5ex\hbox{$\sim$}}}\ht2=\grdimen\dp2=0pt
\def\gax{\mathrel{\copy\gaxbox}}
\def\lax{\mathrel{\copy\laxbox}}
\def\boxit#1    {\vbox{\hrule\hbox{\vrule\kern3pt
                  \vbox{\kern3pt#1\kern3pt}\kern3pt\vrule}\hrule}}
\def\h      {\ifmmode{^{\rm h}}\else{$^{\rm h}$}\fi}
\def\m      {\ifmmode{^{\rm m}}\else{$^{\rm m}$}\fi}
\def\s      {\ifmmode{^{\rm s}}\else{$^{\rm s}$}\fi}
\def\decas    {\ifmmode{{\rlap.}{''}}\else{${\rlap.}{''}$}\fi}
\def\mum     {\ifmmode{\mu{\rm m}}\else{$\mu{\rm m}$}\fi}
\def\s      {\ifmmode{^{\rm s}}\else{$^{\rm s}$}\fi}
\def\deg      {\ifmmode{^{\circ}}\else{$^{\circ}$}\fi}
\def\as     {\ifmmode {\rlap.}$\,$''$\,$\! \else ${\rlap.}$\,$''$\,$\!$\fi}
\def\decsec  {\ifmmode {\rlap.}$\,$^{s}$\,$\! \else ${\rlap.}$\,$^{s}$\,$\!$\fi}\def\decs  {\ifmmode {\rlap.}$\,$^{s}$\,$\! \else ${\rlap.}$\,$^{s}$\,$\!$\fi}

\def\kms    {\ifmmode{{\rm km~s}^{-1}}\else{km~s$^{-1}$}\fi}

\def\ccm    {cm$^{-3}$}

\def\Lsun   {$L_{\odot}$}

\def\Mspy   {\ifmmode {M_{\odot} {\rm yr}^{-1}} \else $M_{\odot}$~yr$^{-1}$\fi}
\def\Mdot   {\ifmmode {\dot M} \else $\dot M$\fi}
\def\mhd    {\ifmmode {n_{{\rm H}_2}} \else $n_{{\rm H}_2}$\fi}
\def\mhcd   {\ifmmode {N_{{\rm H}_2}} \else $N_{{\rm H}_2}$\fi}

\def\El      {\ifmmode{E_{\ell}}\else{$E_{\ell}$}\fi}
\def\beam    {\ifmmode{\theta_{\rm B}}\else{$\theta_{\rm B}$}\fi}
\def\mjyb   {\ifmmode {{\rm mJy~beam}^{-1}} \else{mJy~beam$^{-1}$}\fi}
\def\mujyb   {\ifmmode {\mu{\rm Jy~beam}^{-1}} \else{$\mu$Jy~beam$^{-1}$}\fi}
\def\Trot   {\ifmmode{T_{\rm rot}}\else$T_{\rm rot}$\fi}    
    
\def\Teff   {\ifmmode{T_{\rm eff}}\else$T_{\rm eff}$\fi}

\def\ITRS   {\ifmmode{\smallint {\rm T}_{R}^{*}dv}\else{$\smallint 
{\rm T}_{R}^{*}dv$}\fi}
\def\ITRS   {\ifmmode{\smallint {\rm T}_{R}^{*}dv}\else{$\smallint 
{\rm T}_{R}^{*}dv$}\fi}
\def\ITAS   {\ifmmode{\smallint {\rm T}_{A}^{*}dv}\else{$\smallint 
{\rm T}_{A}^{*}dv$}\fi}

\def\hzo        {H$_2$O}

\def\nuteo {$\nu_2$=1}
\def\nuteo {\mbox{$\nu_2$=1}}
\def\kbw   {\hbox{$6_{16}-5_{23}$}}

\def\HHOOET {\hbox{$3_{13}-2_{20}$}}          
\def\lefttitle#1  {\noindent \hangindent=18.0pt \hangafter=1 {#1} \par}
\def\vol#1  {{\bf {#1}{\rm,}\ }}
\font\tenssb=cmssbx10
\font\tenbf=cmbx10
\font\sevenbf=cmbx8
\font\fivebf=cmbx6
\def\unetdemi    {\smallskipamount=6pt plus2pt minus2pt
                  \medskipamount=12pt plus4pt minus4pt
                  \bigskipamount=24pt plus8pt minus8pt
                  \normalbaselineskip=16pt plus0pt minus0pt
                  \normallineskip=2pt
                  \normallineskiplimit=0pt
                  \jot=6pt
                  {\def\smallskip {\vskip\smallskipamount}}
                  {\def\medskip   {\vskip\medskipamount}}
                  {\def\bigskip   {\vskip\bigskipamount}}
                  {\setbox\strutbox=\hbox{\vrule 
                    height17.0pt depth7.0pt width 0pt}}
                  \parskip 12.0pt
                  \normalbaselines}
\def\smallerspace {\smallskipamount=3pt plus0pt minus0pt
                  \medskipamount=6pt plus0pt minus0pt
                  \bigskipamount=10.5pt plus0pt minus0pt
                  \normalbaselineskip=10.5pt plus0pt minus0pt
                  \normallineskip=1pt
                  \normallineskiplimit=0pt
                  \jot=3pt
                  {\def\smallskip {\vskip\smallskipamount}}
                  {\def\medskip   {\vskip\medskipamount}}
                  {\def\bigskip   {\vskip\bigskipamount}}
                  {\setbox\strutbox=\hbox{\vrule 
                    height8.5pt depth3.5pt width 0pt}}
                  \parskip 0pt
                  \normalbaselines}
\def\memospace    {\smallskipamount=4pt plus1pt minus1pt
                  \medskipamount=6pt plus2pt minus2pt
                  \bigskipamount=14pt plus6pt minus6pt
                  \normalbaselineskip=14pt plus0pt minus0pt
                  \normallineskip=1pt
                  \normallineskiplimit=0pt
                  \jot=4pt
                  {\def\smallskip {\vskip\smallskipamount}}
                  {\def\medskip   {\vskip\medskipamount}}
                  {\def\bigskip   {\vskip\bigskipamount}}
                  {\setbox\strutbox=\hbox{\vrule 
                    height17.0pt depth7.0pt width 0pt}}
                  \parskip 2.0pt
                  \normalbaselines}
\def\memowidespace    {\smallskipamount=5pt plus1pt minus1pt
                  \medskipamount=7.5pt plus2pt minus2pt
                  \bigskipamount=17.5pt plus6pt minus6pt
                  \normalbaselineskip=17.0pt plus0pt minus0pt
                  \normallineskip=1.25pt
                  \normallineskiplimit=0pt
                  \jot=5pt
                  {\def\smallskip {\vskip\smallskipamount}}
                  {\def\medskip   {\vskip\medskipamount}}
                  {\def\bigskip   {\vskip\bigskipamount}}
                  {\setbox\strutbox=\hbox{\vrule 
                    height21.25pt depth8.75pt width 0pt}}
                  \parskip 2.5pt
                  \normalbaselines}
\usepackage{graphicx}
\usepackage{natbib}
\usepackage{txfonts}
\usepackage{ulem}
      \def\new#1 {{\bf #1 }}
      \def\cut#1 {\sout{#1} }

\begin{document}

\title{A multi-transition submillimeter water maser study of evolved stars --
detection of a new line near 475 GHz}
\author{K. M. Menten
\inst{1}
\and
A. Lundgren
\inst{2}
\and
A. Belloche
\inst{1}
\and
S. Thorwirth
\inst{1}
\and
M. J. Reid
\inst{3}
}


\offprints{K. M. Menten}

\institute{Max-Planck-Institut f\"ur Radioastronomie,
Auf dem H\"ugel 69, D-53121 Bonn, Germany
\email{kmenten, belloche, sthorwirth@mpifr-bonn.mpg.de}
\and
ESO, Casilla 19001, Santiago 19, Chile
\email{alundgre@eso.org}
\and
Harvard-Smithsonian Center for Astrophysics
60 Garden Street
Cambridge, MA 02138, USA
\email{reid@cfa.harvard.edu}
}

\date{Received / Accepted}
\titlerunning{Submillimeter water maser lines in evolved stars}

\authorrunning{Menten et al.}

 \abstract
   {Maser emission from the  \hzo\ molecule probes the warm,   inner circumstellar envelopes of oxygen-rich red giant and supergiant stars. Multi-maser transition studies can be used to put constraints on the density and    temperature of the emission regions.}
   {A number of known \hzo\ maser lines were observed toward the long period variables R Leo and W Hya and the red supergiant VY CMa. A search for a new, not yet detected line near 475 GHz was conducted toward these stars.}
   {The Atacama Pathfinder Experiment telescope was used for a multi-transition observational study of submillimeter \hzo\ lines.}
   {The $5_{33} - 4_{40}$ transition near 475 GHz  was clearly detected toward VY CMa and W Hya. Many other \hzo\ lines were detected toward all three target stars. Relative line intensity ratios and velocity widths were found to vary significantly from star to star.
}
   {Maser action is observed in all but one line for which it was theoretically predicted. In contrast, one of the strongest maser lines, in R Leo by far \textit{the} strongest, the 437 GHz $7_{53}-6_{60}$  transition, is not predicted to be inverted. Some other qualitative predictions of the model calculations are at variance with our observations. Plausible reasons for this are discussed. Based on our findings for W Hya and VY CMa, we find evidence that the \hzo\ masers in the AGB star W Hya arise from the regular circumstellar outflow, while shock excitation in a high velocity flow seems to be required to excite masers far from the red supergiant VY CMa.}


\maketitle

\section{\label{intro}Introduction}
Water (\hzo) is one of the most abundant molecules in the atmospheres and envelopes of oxygen-rich red giant and red supergiant stars
\citep[see, e.g., ][]{Tsuji1964, Ohnaka2004}.
Often very intense \hzo\ maser emission in the 22.2 GHz\
$J_{K_{a}K_{c}}$\footnote{Energy levels of the asymmetric rotor molecule \hzo\ are denoted $J_{K{\rm a},K_{\rm c}}$ where $J$ is the total rotational quantum number and $K_a$ and $K_c$ are its projections on the orthogonal symmetry axes of the limiting prolate and oblate symmetric rotor, respectively. For ortho transitions $K_{a}+K_{c}$ is odd and for para transitions $K_{a}+K_{c}$ is even.} $= 6_{16} - 5_{23}$ transition has been found toward many hundreds of low- and high-mass star-forming regions and circumstellar envelopes of Mira and semi-regular long-period variable stars (LPVs), as well toward a few luminous
red supergiants \citep[see, e.g., ][]{Valdettaro_etal2001}. Since the $6_{16}$ rotational energy level is high (643 K) above the ground-state and its transition probability is rather small (because of its low frequency) this line is readily observable
even from sea level.

\begin{figure}[t]
\begin{center}
\includegraphics[width=7.5cm,angle=0]{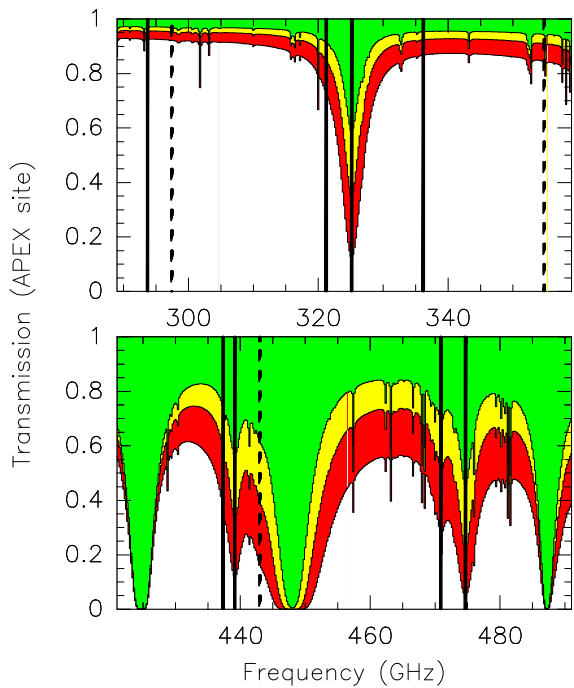}
\caption{\label{transmission}Calculated atmospheric \textit{zenith} transmission for the site of the APEX
telescope for three values of the precipitable water vapor column: 0.2 mm (upper curve), 0.4 mm (middle curve) and 0.8 mm (lower curve). The upper panel shows the frequency range around \hzo\ lines that have been observed with APEX in the $870~\mu$m window, while the lower panel shows the frequency range around lines in the $650~\mu$m window. Positions of detected \hzo\ lines are marked as vertical lines, \hzo\ lines that were searched, but not detected as dashed lines. Observations of the 293, 297, and 336 GHz lines from the \nuteo\ state have been reported by \citet{Menten_etal2006}, observations of all the other lines in the present paper. Note that, both, FLASH and APEX 2a are double sideband receivers with intermediate frequency bands centered at 3 and 6 GHz, respectively. Signal and image sideband are thus separated by 6 and 12 GHz, respectively, and frequently have significantly different transmission.}
\end{center}
\end{figure}

In addition to the 22.2 GHz line, maser emission from a number of other ortho and para \hzo\ (sub)millimeter-wavelength transitions from within the vibrational ground state has been detected toward star-forming regions and red giant and supergiant stars \citep{Menten_etal1990a, Menten_etal1990b, Cernicharo1990, Melnick_etal1993, Yates_etal1995, Liljestrom_etal2002}. All of these lines, except for the 183 GHz \HHOOET\ transition \citep{Cernicharo1990}, were also observed in the present study and are listed in Table \ref{lines}\footnote{Frequencies and energy values are taken from the JPL catalog: http://spec.jpl.nasa.gov/ftp/pub/catalog/c018003.cat}.
Because water vapor in the Earth's atmosphere absorbs heavily in many of these (and other) \hzo\ lines, they are only observable from high, dry sites and several only during the small percentage of time when the atmosphere's \hzo\ vapor content is minimal over the Atacama Pathfinder Experiment's 5100m high site on the Llano de Chajnantor; see Fig. \ref{transmission}. Note that, in particular, the newly discovered 475 GHz line must be amongst the submillimeter lines with the poorest transmission detected from the ground so far.

 Toward the hot, innermost envelopes of red giants and supergiants rotational lines from within the lowest vibrationally excited state, the bending mode (\nuteo), have also been found \citep{MentenMelnick1989, MentenYoung1995, Liljestrom_etal2002}; see \cite{Hunter_etal2007} for a compilation and interferometric measurements. Given their excitation conditions, these lines must arise from within a few stellar radii of the photosphere \citep{Menten_etal2006}.

 \citet[][hereafter NM91]{NeufeldMelnick1991} performed statistical equilibrium calculations to analyze the excitation of \hzo\ masers in the vibrational ground state and outlined parameter regions (of, essentially, temperature and density) for which the energy levels of certain transitions become inverted.
  They predicted inversion in all the lines listed in Table \ref{lines}, except for the 437 and 443 GHz lines. Maser action has indeed been observed in all but one of the predicted lines toward star-forming regions and/or red giant stars.
  (Maser emission in the 475 GHz
  $5_{33} - 4_{40}$ is for the first time reported here.) In particular,
  NM91 find the very widespread 22.2 GHz \kbw\ transition to be the most strongly inverted one over a very wide range of physical conditions. These authors \textit{do not} predict maser emission for the 437 GHz $7_{53}-6_{60}$ transition first detected by
  \citet{Melnick_etal1993} in evolved stars and also observed by us, but \textit{do} predict maser action in the extremely high excitation $17_{4,13} - 16_{7,10}$ line near 355 GHz line, which we searched for, but did not detect.

Using the Atacama Pathfinder Experiment (APEX\footnote{This publication is based on data acquired with the Atacama Pathfinder Experiment (APEX). APEX is a collaboration between the Max-Planck-Institut f\"ur Radioastronomie, the European Southern Observatory, and the Onsala Space Observatory.}) 12m telescope we conducted a multi-transition submillimeter \hzo\ maser study, observing the lines listed in Table \ref{lines}.
Our dataset is complemented by spectra of the 22.2 GHz \hzo\ line taken with the Effelsberg 100m telescope close in time to our APEX observing run.
Our targets were two asymptotic giant branch (AGB) objects, the Mira star R Leo and the semiregular (SRa) variable W Hya, and the red supergiant (RSG) VY CMa. The former two are typical, nearby, LPVs with ``reprocessed'' \citep{Knapp_etal2003} Hipparcos parallax distances, $D$, of
$82^{+11}_{-9}$ pc and $78\pm6$ pc, respectively.
Their mass loss rates\footnote{In the following all quoted mass-loss rates and luminosities are  values taken from the quoted references, but corrected for the revised distances quoted in the text.}, \Mdot, derived from CO observations are several times $10^{-7}$ \Mspy\ \citep{Young1995}, although for both objects significantly different values have been derived \citep[for R Leo, see][and \S\ref{differentnature} for W Hya]{MentenMelnick1991}.

With a luminosity, $L$, of $2\,10^5$\,\Lsun\ \citep{Sopka1985} and a mass-loss rate of  1--$2\,10^{-4}$\,\Mspy\ \citep{Danchi1994} the peculiar red supergiant VY\,CMa is 
a remarkable object by any standard and the most prolific stellar (OH, \hzo, and SiO) maser source known. Its distance is usually assumed to be 1.5 kpc \citep{LadaReid1978}. However, a recent VLBI trigonometric parallax of VY CMa's SiO masers yields a distance $\approx20$\% closer (Reid \&\ Menten, in preparation).  In the following we shall assume that the distance is 1.1 kpc.

Our observations are summarized in \S\ref{obs}. In \S\ref{results} we describe our results. In particular, we report the detection of a new maser line, the  $5_{33} - 4_{40}$  transition near 475 GHz. We discuss the observed spectra, their similarities, but also very pronounced
differences in the relative luminosities  and the spectral appearances of the lines for each object and also among objects. In \S\ref{discussion} our observational results are compared with predictions of calculations that aim at modeling \hzo\ maser emission. We present a summary in \S\ref{summary}.

\begin{figure*}[t]
\begin{center}
\includegraphics[width=13.5cm,angle=0]{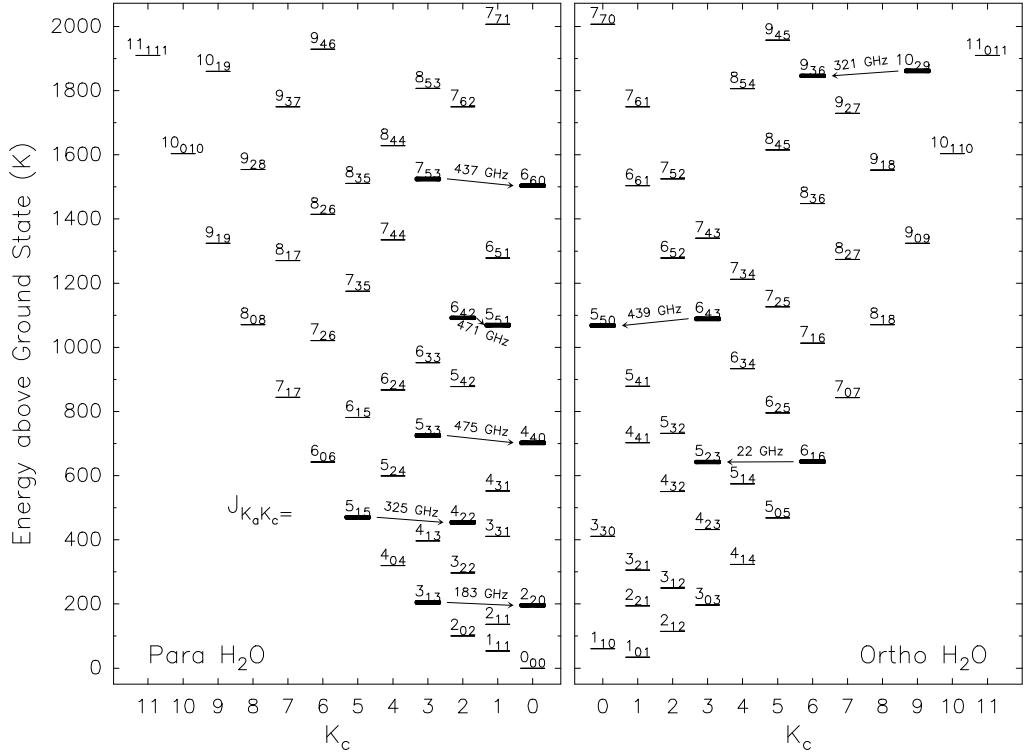}
\caption{\label{energylevels}Energy level diagram of para \hzo\ (left) and ortho \hzo\ (right).
Upper and lower levels of known maser lines appear in bold, are connected by arrows and have the rounded transition frequency (in GHz)
indicated. Data for all these lines (except for the 183 GHz line) are reported in the present
paper.}
\end{center}
\end{figure*}

\section{\label{obs}Observations and data reduction}
Our observations  were made between 2006 June 7 and 19 under generally excellent weather conditions with the 12\,m Atacama Pathfinder Experiment telescope \citep{Gusten_etal2006}.

All lines considered here are listed in Table\,\ref{lines}. The 434, 437, 439, 443, 471, and 475 GHz lines (see Table \ref{lines}) were observed with the First Light APEX Submillimeter Heterodyne instrument 
\citep[FLASH, ][]{Heyminck_etal2006} while the APEX 2a facility receiver \citep{Risacher_etal2006} was used to observe the 321, 325, and 355 GHz lines as well as the $J = 3 - 2$ line of CO. Calibration was obtained using the chopper wheel technique under consideration of the very different atmospheric opacities in the signal and image sidebands of the employed double sideband receivers (see Fig. \ref{transmission}). The radiation was analyzed with the MPIfR Fast Fourier Transform spectrometer, which provides 16384 frequency channels  over the 1\,GHz intermediate frequency bandwidth \citep{Klein_etal2006}. To increase the signal to noise ratio, the spectra were smoothed to effective velocity resolutions appropriate for the measured linewidths, typically $\sim0.5$--1 \kms. To check the telescope pointing, the receiver was tuned to the 437\,GHz \hzo\  line, which showed strong emission in all of our three sources, and five point crosses centered on the stellar position with half beamwidth offsets in elevation and azimuth were measured. Pointing corrections were derived from the latter measurements. The pointing
was found to be accurate to within $\approx3''$, acceptable
given the FWHM beam size, $\theta_{\rm B}$, which is $20''$ FWHM at 321 GHz and $13''$ at 475 GHz.

\begin{figure*}[t!]
\begin{center}
\includegraphics[width=13.5cm,angle=0]{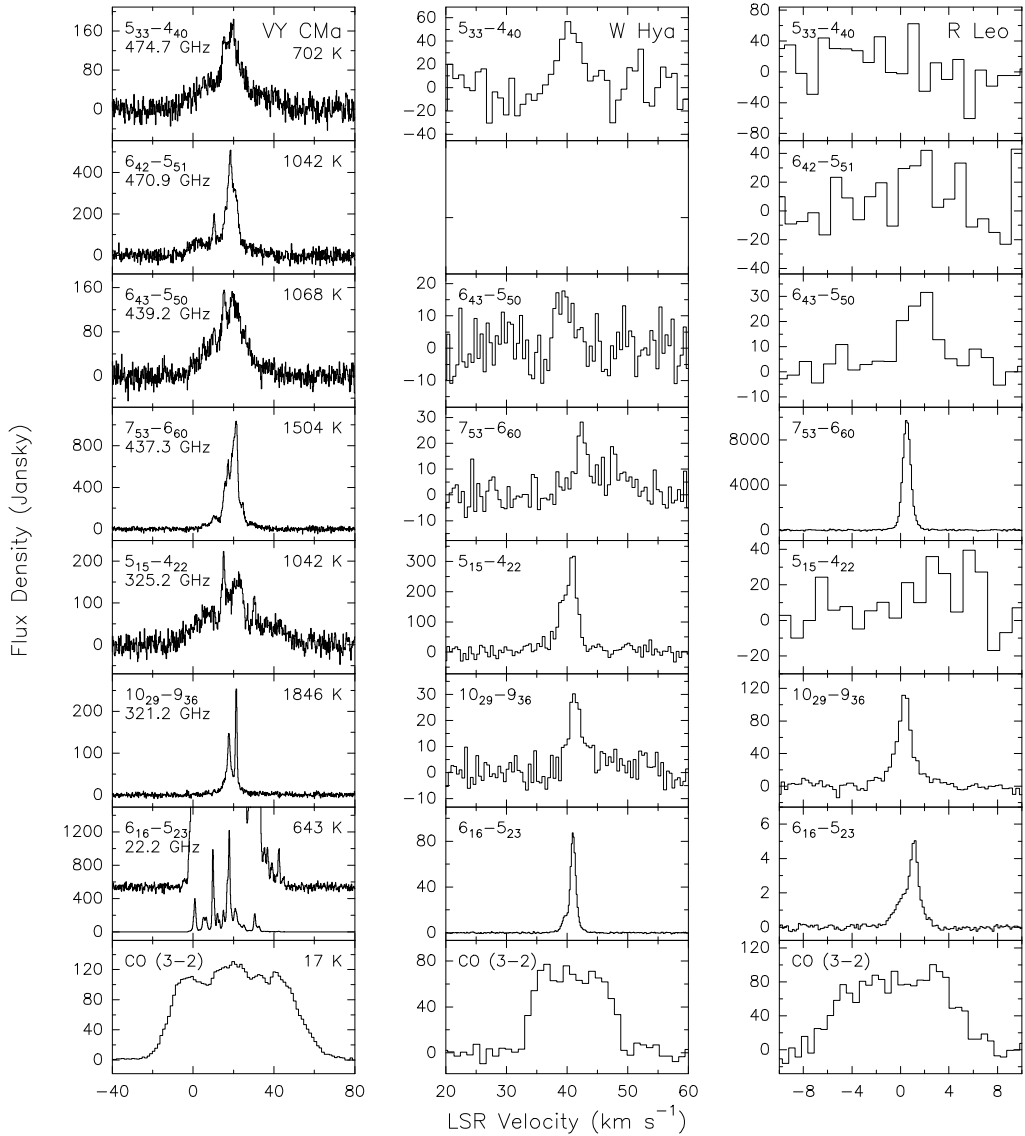}
\caption{\label{observedspectra}Top to second panel from
bottom: \hzo\ spectra observed
toward VY CMa, W Hya, and R Leo (left, middle, and right column,
respectively). The different \hzo\ lines are labeled. In the VY CMa panels also the approximate rest frequencies and energies above ground are listed. To bring out weak features and better show the full extent of its emission, the
spectrum of the 22.2 GHz line is also shown magnified by a factor of 100.
The bottom panels shows the 345.8 GHz $J = 3 - 2$ line of CO. Note that some of the spectra have very poor signal-to-noise ratios, which may lead to ``broadening'' of the feature; cf. the 439 and 475 GHz lines in W Hya. Similarly, it is difficult to tell how much of broad emission seen in some of the lines from VY CMa is due to blended narrow features. Also, the ``bump'' at \textit{v} $> 45$~\kms\ in the 437 GHz spectrum of W Hya is almost certainly a baseline artefact. These facts should be kept in mind in the discussion in \S\ref{nature}.}
\end{center}
\end{figure*}

Additional data had been taken earlier toward VY CMa for the 355 GHz line (see Table \ref{lines}) for which maser action had been predicted by excitation modeling (see \ref{predictions}). These observations were made in 2005 July/August.

In Table \ref{lineresults} we present our line intensities in a flux density scale (i.e., in Janskies) assuming the aperture efficiencies observationally determined by \citet{Gusten_etal2006} for the respective frequency ranges.

Observations of the 22.2 GHz \kbw\ transition were made with the Effelsberg 100m telescope on 2006 June 27, i.e., ca. 1--2 weeks after the submillimeter observations. The line was detected with the facility high electron mobility transistor receiver
and autocorrelator backend. The data were corrected for atmospheric opacity and variations of the telescope's gain curve with elevation.

Fig. \ref{energylevels} shows our observed \hzo\ lines on energy level diagrams of para and ortho water.

\begin{table}[t]
\begin{center}
\caption{\label{lines}Ground-state water maser lines observed with APEX.}
\begin{tabular}{clr}
 \hline \hline
\hzo               &Frequency     &$E_\ell/k$\d\\
$J'_{K'_{a}K'_{c}} - J''_{K''_{a}K''_{c}}$
= & \d\d(MHz)              &   (K)\d \\
 $6_{16} - 5_{23}     $ & \d22235.08 & 642.5\\
$10_{29} - 9_{36}     $ & 321225.64 &1845.9\\
 $5_{15} - 4_{22}     $ & 325152.92 & 454.4\\
$17_{4,13} - 16_{7,10}$ & 354808.9  &5764.3\\
 $7_{53} - 6_{60}$ & 437346.67 & 1503.7\\
 $6_{43} - 5_{50}$ & 439150.81 & 1067.7\\
 $7_{52} - 6_{61}$ & 443018.30 & 1503.7\\
 $6_{42} - 5_{51}$ & 470888.95 & 1041.8 \\
 $5_{33} - 4_{40}$ & 474689.13 &  702.3 \\
\noalign{\smallskip}
 \hline
 \noalign{\smallskip}
 \end{tabular}
\end{center}
Columns are (from left to right) quantum numbers of upper and lower state, frequency and energy above ground of lower state in Kelvins; $k$ is the Boltzmann constant.
Frequency values, taken from the JPL catalog,
have formal uncertainties of order 50 kHz. More accurate values from a fit to the \hzo\ spectrum have been presented by \citet{Chen2000}. The difference between their values and  the ones used by us is typically of order 20 kHz, corresponding to less than 0.02 \kms, which is smaller than the uncertainties in our velocity determinations.
\end{table}

\begin{table}[tb]
\begin{center}
\caption{\label{lineresults}Results of APEX observations}
\begin{tabular}{rlcrlr}
 \hline \hline
Line     & $\int S$~dv & v-range & $S_{\rm p}$ & \d\d v$_{\rm p}$&$L_\nu$\dd\\
         & (Jy km~s$^{-1})$& \d\d (\kms)  & (Jy)& (\kms)  & (s$^{-1}$)\d\\
\multicolumn{6}{c}{------------------------------------ R Leo ------------------------------------}\\
 22   & \dd \d  7.0(0.1)      & [$-$1.3,+3.2]& 5 & \d  1.1&$ 3~10^{40}$\\
321   & \d 186(6)          & [$-$2.7,+2.5] & 111 & \d  0.4&$ 6~10^{41}$\\
325   & \dd (159)            &   --          & (105)& \d -- &$<4~10^{40}$ \\
355   & \dd \dd --             &   --  & \d --  & \d --   & --\dd\\
437   &8452(19)         & [$-$1.4,+2.4] &8074 & \d  0.53  &$ 3~10^{43}$\\
439   & \dd  87(7)          & [$-$0.4,+3.1] &  35 & \d 2  &$ 4~10^{41}$\\
443   & \dd(90)            &    --         & (63)& \d --  &$<4~10^{41}$\\
471   & \dd 95(24)         & [$-$0.2,+2.7] &  59 & \d  2  &$ 4~10^{41}$\\
475   & \dd (171)             &  --           & (108)& \d --&$<7~10^{41}$\\
\multicolumn{6}{c}{----------------------------------- W Hya -------------------------------------}\\
 22   & \d 116.0(0.4)      &  [+37.7,+43.9]&  87 & 41.0   &$ 4~10^{41}$\\
321   & \dd 90(13)         &    [+39,+45]  &  29 & 41     &$ 3~10^{41}$\\
325   & \d 871(55)         &    [+37,+43]  & 339 & 41     &$ 3~10^{42}$\\
355   & \dd \dd --             &   --      & --  & \d --  & --\dd\\
437   & \dd 50(8)          &    [+37,+43]  &  29 & 42     & $2~10^{41}$\\
439   & \dd 56(12)         &    [+37,+43]  &  18 & 39     & $2~10^{41}$\\
443   & \dd \dd --             &  --      & --  & \d --   & --\dd\\
471   & \dd \dd --             &   --     & --  & \d --   & --\dd\\
475   & \d 213(30)         &    [+38,+44]  &  54 & 40     & $8~10^{41}$ \\
\multicolumn{6}{c}{----------------------------------- VY CMa ---------------------------------}\\
 22   &5993(1)          & [$-$5.1,+45.4]&4064 & 17.9      & $4~10^{45}$ \\
321   & \d 859(21)         &     [+5,+42]  & 238 & 21.4   & $6~10^{44}$ \\
325   &3836(132)        &   [$-7$,+54]  & 220 & 15        & $3~10^{45}$ \\
355   & \dd \d (15)             &   --     & (3) & \d --   &$<1~10^{43}$ \\
437   &6913 (49)        &[$-$19.1,+35.5]&1015 & 21.5      & $5~10^{45}$ \\
439   &2425(38)         &   [$-$2,+41]  & 153 & 15        & $2~10^{45}$ \\
443   & \dd (270)            &    --         &(54) & \d -- &$<2~10^{44}$ \\
471   &4171 (65)        & [$-$5.5,+40.2]& 509 & 18.4      & $3~10^{45}$ \\
475   &2798 (55)        &   [$-$7,+48]  & 171 & 19        & $2~10^{45}$ \\
\noalign{\smallskip}
 \hline
 \noalign{\smallskip}
 \end{tabular}
\end{center}
\footnotesize{Columns are (from left to right) line ID (i.e. frequency in GHz; see Table \ref{lines}), integrated flux density, velocity range covered by line (FWZP), peak flux density of strongest emission, velocity of strongest emission
and isotropic photon luminosity. Numbers in parentheses following integrated flux densities are formal errors derived from Gaussian fitting and/or noise level analysis. The overall calibration has an estimated uncertainty of 20\%. For non-detections, the quoted flux density (in parentheses) is three times the rms noise value of the spectrum smoothed to $\approx0.5$ \kms\ resolution and the quoted value of the integrated flux density has been calculated assuming a width equal to that of the \hzo\ line found in the respective source that covers the widest velocity range. To calculate isotropic luminosities, distances of 82, 78, and 1100 pc have been assumed for R Leo, W Hya, and VY CMa, respectively. For upper limits of the latter three times the rms noise in the integrated flux density quoted in column 2 has been assumed.}


\end{table}
%
%
%
%

\section{\label{results}Results}
\subsection{\label{nature}General properties of the observed emission -- luminosities}
\begin{figure}[b]
\begin{center}
\includegraphics[width=6.5cm,angle=0]{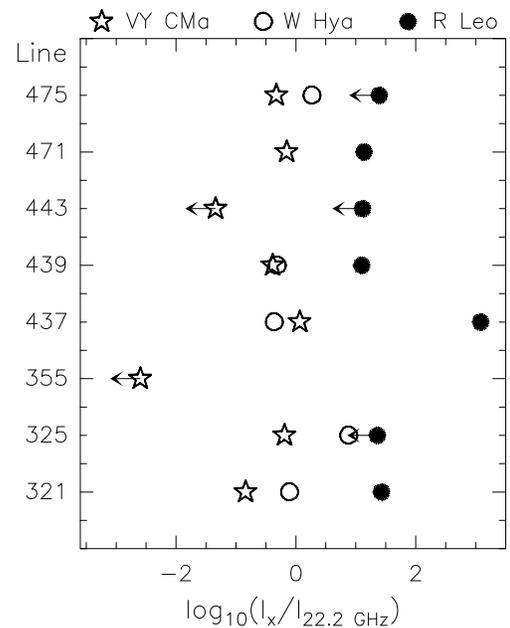}
\caption{\label{lineratios} Ratios of the integrated submillimeter \hzo\ maser line flux densities (marked by their frequency rounded to GHz) to the integrated line flux density of the 22.2 GHz line for our three program stars, R Leo (filled circles), W Hya (open circles), and VY CMa (star symbols). For upper limits, 3 times the rms levels given in Table \ref{lineresults} were used. Note that I$_{\rm x}$ for a line is proportional of its photon luminosity.}
\end{center}
\end{figure}

For W Hya, the observed isotropic photon luminosities of the detected \hzo\ maser lines are all within one order of magnitude, ranging from  $3~10^{41}$ to $3~10^{42}$~s$^{-1}$ (Table \ref{lines} and Fig. \ref{lineratios}). This is also true for VY CMa, but here the much higher luminosities range from
$6~10^{44}$ to $5~10^{45}$~s$^{-1}$.
The situation is different with R Leo. Here the  321 and 439 and the marginally detected 471 GHz line have comparable luminosities of $8~10^{41}$, $4~10^{41}$~s$^{-1}$, and $4~10^{41}$~s$^{-1}$, respectively. However,  the 22.2 GHz line is anomalously weak ($3~10^{40}$~s$^{-1}$) and the 437 GHz line extremely bright ($3~10^{43}$~s$^{-1}$).

We emphasize in particular that for W Hya and VY CMa the luminosity of the 22.2 GHz lines is comparable to that of other lines and that for R Leo it is much lower. This is important for the discussion of maser excitation (see \S\ref{excitation}).

\subsection{\label{spectralappearance}Spectral appearance -- clues to the emission regions}
As can be seen from Table \ref{lines} and Fig. \ref{energylevels}, the lines observed by us arise from energy levels that span a very wide range of energies above the ground state, with the lower level energy, $E_\ell/k$, ranging from 454 to 1846 K. The highest excitation lines (437 GHz/1505 K and 321 GHz/1846 K) cover a smaller velocity extent than lower excitation lines. One might expect this if they are formed in a hot region closer to the star with an outflow velocity that is smaller than further out from where the lower excitation lines are arising. However, there is no clear monotonous
trend from larger to smaller linewidth with increasing energy above ground.

All the \hzo\ lines observed toward R Leo and W Hya (Fig. \ref{observedspectra}) show a single
feature or a blend of features centered on or within a few \kms\ of their stellar
velocities determined by the CO line centroids. The features are asymmetric, except for the 437 GHz line toward R Leo, which is nearly Gaussian with a FWHM of 0.78 \kms. In contrast, for VY CMa essentially every line has a different shape.

For both R Leo and W Hya, the FWZP width of the widest \hzo\ line is only $\approx 20$\%\ of the CO line's width. In contrast, for VY CMa, some lines cover $\approx 50$ \kms, which is about half the velocity range of the CO line. This indicates that for all three stars 
the \hzo\ maser emission arises from a part of the envelope in which the outflow has not yet reached its terminal velocity (i.e., half the FWZP linewidth of the CO line shown in the bottom panels of
Fig. \ref{observedspectra}).  This is plausible, given the small size emission region  of the 22.2 GHz emission, which for W Hya was found to be a ring with a diameter of $0\as3$
\citep[24 AU, ][]{ReidMenten1990} and for VY CMa has a size of $0\as7\times0\as4$ that translates into a much larger linear size of $770\times440$ AU \citep{RichardsCohen1998}. The latter is still much smaller than size of the CO emission region of several thousand AU
\citep{Muller_etal2007}.

In contrast, a spectrum of the \textit{thermally excited} 557 GHz $1_{10}-1_{01}$ ortho ground-state \hzo\ line observed toward W Hya with the Odin satellite has a width (and centroid velocity) that is consistent with the CO values \citep{Justtanont_etal2005}.

For VY CMa the situation is more complicated. Toward this source,
various (but not all) thermal and maser lines from SiO show velocities equal or even exceeding the values seen for CO,
indicating high velocity motions that do not fit into a continuous outflow
picture \citep{Cernicharo_etal1997,Herpin_etal1998}. Given their excitation requirements, the vibrationally excited SiO maser lines must arise from the closest vicinity of the stars. This indicates that high velocity components are already present on the smallest scales and receive their peculiar high velocities, sometimes exceeding the terminal velocity, by other processes
than the ``regular'' outflow, which only can commence further out, once dust starts forming.
A possible mechanism may be related to shocks driven into the atmosphere and inner envelope by the arrival of giant convection cells to the star's surface layers \citep{Schwarzschild1975} and, e.g., \citet{JosselinPlez2007}.
Puzzlingly, we see no sign of these extreme velocities in any of the water lines, neither in the ground-state lines nor in vibrationally excited \hzo\ lines, which all  cover much smaller velocity ranges than high velocity SiO maser lines, although the excitation requirements of vibrationally excited \hzo\ and SiO lines are similar \citep{MentenMelnick1989,MentenYoung1995,Menten_etal2006}. These lines do, in fact, cover velocity ranges comparable to those of the 321 and 437 GHz \hzo\ lines observed by us. Indeed, under plausible physical conditions for the regions from which the vibrationally lines arise, $n({\rm H}_2) = 2~10^9$ \ccm\ and $T = 1000$ K \citep{Menten_etal2006}, one also finds inversion of these ground state lines (see \S\ref{excitation}).

A spectrum of the thermally excited $1_{10}-1_{01}$ ortho ground-state \hzo\ line measured with Submillimeter Wave Astronomy Satellite (SWAS) has been published by \citet{HarwitBergin2002}. They  find a FWZP value of 50 \kms, similar to that we find for the wider of the maser lines. It is, in contrast to the case of W Hya, significantly smaller than that of CO lines (see above).

The variety of \hzo\ line shapes toward VY CMa ranges from that of the 22.2 GHz spectrum, which entirely consists of narrow spikes, to spectra in which most or an appreciable portion of the line flux arises from a broad (up to 50 \kms\ wide) pedestal feature on which a few spikes are superposed (in the 325, 439, 471, and 475 GHz line). The 321 and 437 GHz spectra are dominated by a few, strong narrow features with a little pronounced broad pedestal. The former covers a much narrower range than the other lines, just $\pm$ a few \kms\ around the stellar velocity.

That the kinematics around VY CMa are indeed complex also on scales much larger than the masing regions is demonstrated by
interferometric CO observations which are interpreted by \citet{Muller_etal2007} in terms of spherically symmetric outflow together with a high velocity
bipolar flow.

\section{\label{discussion} Discussion}
\subsection{\label{excitation}Modeling the excitation of water maser lines}
Earlier excitation studies aimed at explaining stellar (and interstellar) \hzo\ maser pumping only considered a limited number of energy levels, both, due to limitations of computational capabilities and because collision rates were not known for excitation to and from higher energy states \citep[see, e.g., ][]{deJong1973, Deguchi1977, CookeElitzur1985}. Nevertheless, all these studies predicted strong inversion of the 22.2 GHz line, which was the only \hzo\ maser line
known at the time. These studies also predicted maser action in a number of
other transitions that were later indeed found to be masing, despite the fact that in some cases the collisional excitation rate coefficients used were vastly different from the ``modern'' values computed by \citet{Palma_etal1988}. The more recent,  comprehensive, study by NM91 considered all 349 states of ortho and para water up to an energy of 7700 K above the ground state and extrapolated the Palma et al. rates to high enough temperatures.

\citet{NeufeldMelnick1990} and NM91 performed statistical equilibrium calculations modeling vibrational ground state \hzo\ excitation and radiative transfer for plane parallel media under the large velocity gradient (LVG) assumption.
\citet{NeufeldMelnick1990} specifically explore conditions under which, both, the the 22.2 GHz and the 321 GHz lines are masing for the case of an O-rich evolved star with a mass-loss rate ($3~10^{-5}$ \Mspy) intermediate between the values of our objects. They find that the pumping for these lines is dominated by collisions and that radiation plays a minor role. The same was also found for other transitions in circumstellar envelopes \citep[][ NM91]{Deguchi1977, CookeElitzur1985}.

NM91 explored larger regions of (essentially) density and temperature parameter space in which  the energy levels of certain transitions become inverted. They calculate  maser emissivities and optical depths as functions of temperature and a variable, $\zeta'$, which is equal to a geometric factor, $G$, divided by the velocity gradient, $d{\rm v}/dr$, times the product of the hydrogen nuclei, $n$, and the water density, $n({\rm H}_2{\rm O})$, i.e., $n\times n({\rm H}_2{\rm O})$ $ = n^2 x({\rm H}_2{\rm O})$, where $x({\rm H}_2{\rm O})$ is the \hzo\ abundance. $\zeta'$ is, thus, given by:

$$\zeta' = G{n^2 x({{\rm H}_2{\rm O})}\over{d{\rm v}/dr}}$$
Assuming a plausible, constant, value for the \hzo\ abundance and the velocity gradient of $10^{-4}$ and $10^{-8}$~cm~s$^{-1}$~cm$^{-1}$, respectively,
makes $\zeta'$ dependent on the H density squared alone.
Measuring the H density in units of $10^9$ \ccm\ in the parametrization  of NM91, values of $log_{10}\zeta' = -1, 0,$ and 1 correspond to H densities of 0.32, 1, and $3.2\times10^9$ \ccm.

\subsection{\label{comparison}Comparison with the APEX observations}
NM91 predicted maser emission in all of the lines
listed in Table \ref{lines}, except for the 437 GHz $7_{53}-6_{60}$ and 443 GHz  $7_{52} - 6_{61}$ transitions in, roughly, this range of values quoted above.
Maser emission in all their predicted maser lines has indeed been observed in star-forming regions and/or red giant stars, except for the 355 GHz $17_{4,13} - 16_{7,10}$ transition. They present opacities and emissivities (which are proportional to isotropic photon luminosities) for assorted lines calculated for two values of the kinetic temperature, 400 K and 1000 K. In particular, they find the very widespread 22.2 GHz \kbw\ transition to be by far the most luminous and the most robust maser line in the sense that it is inverted over the widest range of physical conditions in particular at extreme values of the density. A general trend is that for 1000 K the maximum emissivities in all the lines are reached for a factor of a few higher densities than at 400 K.

In the calculations of NM91, the peak emissivities of all lines but the 22.2 GHz line are similar within a factor of a few (although they occur at different values of $\zeta'$). At, both, 400 and 1000 K, the peak emissivity of the 22.2 GHz line is much higher than that of any of the others, by a factor of $\sim10$ at 400 K and $\sim5$ at 1000 K. The peak emissivity is also reached at an order of magnitude higher value of $\zeta'$ than for other lines. As can be seen from Table \ref{lines} and Fig. \ref{lineratios} in none of our stars does the 22.2 GHz line show the behavior predicted by these calculations, i.e. that it should be much more luminous than all the other lines.  While it is among the strongest lines in VY CMa, it is anomalously weak in R Leo. We note that R Leo showed an even more extreme 321/22.2 GHz line ratio in 1989/1990 \citep{MentenMelnick1991}. In this star, the 22 GHz line is known to undergo extreme and erratic variations changing from not or barely detectable to hundreds of Jy within dozens of days \citep{Rudnitskij1987}.

\subsection{The ''new'' 475 GHz transition}
For the newly discovered 475 GHz $5_{33} - 4_{40}$ transition NM91 predict, as for the 439 and 471 GHz lines, maser action only for their 1000 K calculation.
The presence of maser action in this line is, thus, an important indicator of
a high temperature environment.

\subsection{\label{predictions}Non- and mispredictions}
NM91\textit{do not} predict maser emission for the 437 GHz $7_{53}-6_{60}$ transition, which we find to be by far the strongest line in R Leo and the second strongest in VY CMa. This line has only been found toward evolved stars and not toward SFRs \citep{Melnick_etal1993}.
Pumping by the strong IR field in these objects, possibly via line overlaps, not considered by NM91, may be responsible for its excitation. NM91 \textit{do}, however, predict maser action  in the extremely high excitation (5764.3 K) $17_{4,13} - 16_{7,10}$  transition near 355 GHz high with peak emissivities at $\zeta'$ values 1--2 order of magnitude higher than values for which other lines show their peak emissivities. For constant water abundance and velocity gradient this translates into H$_2$ densities of order $10^{10-11}$ cm$^{-3}$, which is 1--2 orders of magnitudes  higher than the values over which most of the other lines show maser action. Our non-detection of this line (see Table \ref{lineresults} and \S\ref{comparison}) seems to indicates that under such extreme conditions the necessary requirements for producing detectable maser emission may not be met.

\subsection{A note of caution}
Guided by the described model calculations, observations of combinations of maser lines in principle should put constraints on the physical parameters of the masing regions. One naive hope might be: The more lines one finds masing the tighter the physical parameters of the masing region can be constrained. One might even hope that the observed relative luminosities of the lines might provide yet tighter constraints.

Unfortunately at present, in the absence of more realistic models of the maser regions such expectations are most likely overly optimistic for several reasons: Comparing the emissivities or their ratios  predicted by the generic calculations of NM91 for a \textit{plane-parallel(!)} medium with measured data probably does not make much sense. Moreover, an inherent assumption in the values calculated by NM91 is that all the lines are saturated throughout the region in which they are inverted. There is good observational evidence that this is not true for the 22.2 GHz line around W Hya \citep{ReidMenten1990} although this issue is unclear (and difficult to address) for other transitions and other stars.

Another caveat when using line ratios as diagnostics is time variability. While all our APEX and Effelsberg data were taken within 1--2 weeks, at least the 22.2 GHz line in some sources (including R Leo) is known to show week-to-week variability (see \S\ref{comparison}). Information on variability on such small time scales is not available for the submillimeter lines, but presumably all of them are variable. This has been directly shown from a comparison of line profiles taken at different epochs 1--2 months apart for the 321 and
325 (and 22.2) GHz lines \citep{MentenMelnick1991,YatesCohen1996}

Furthermore, to use multi-line data as diagnostics, an inherent assumption is that all lines arise from the same region (whose physical parameters are to be constrained). However, for VY CMa, the different shapes of the lines and the different velocity ranges they cover suggest that this is not true and that at least a significant portion of the lines' emission arises from disjoint regions. Clearly, high resolution interferometric observations are necessary to get a clearer picture of this. Given the high expected brightness temperatures, such observations with the Atacama Large Millimeter Array (ALMA) will afford studies of R Leo and W Hya with a resolution better than a stellar radius and a few stellar radii for VY CMa \citep[see \S4.4 of ][]{Menten2000}. All the lines discussed here have frequencies within  ALMA's 275--370 and 385--500 GHz ``First Light'' Receiver bands \citep[``Bands 7 and 8'', ][]{Wilson_etal2005}.
Observations of the 321 and 325 GHz lines can already be made today with the Submillimeter Array \citep{Ho_etal2004}.

It is clear that current models for \hzo\ maser excitation in circumstellar envelopes fail to explain the observational picture. Velocity-resolved observations of a large number of maser and non-maser lines from a wide range of energies above the ground state with the Heterodyne Instrument for the Far Infrared (HIFI) aboard the Herschel satellite to be launched in the near future will soon deliver very tight constraints on \hzo\ excitation. Together with high spatial resolution data for high excitation (mostly maser) lines (see above), which will \textit{only} be attainable with ground-based interferometers  and more sophisticated radiative transfer modeling these observations will lead to an understanding of water, which dominates the thermal balance in these regions.

\subsection{\label{differentnature}Different natures of AGB star and red supergiant water masers?}
Let us come back to the fact, mentioned in \S\ref{spectralappearance}, that toward W Hya, the 22.2 GHz \hzo\ maser emission arises from a ring with a diameter of 24 AU centered on the star. In contrast, toward VY CMa, the maser emission arises from a $770\times440$ AU region and proper motion observations appear to indicate that the masers partake in a bipolar outflow \citep{RichardsCohen1998}, although we notice that this flow would be directed perpendicular to the larger-scale CO outflow imaged by \citet{Muller_etal2007}. Are these distributions consistent with the masers arising from a ''normal'', i.e. steadily expanding circumstellar outflow?

In the following, we investigate whether over the whole of the maser distributions in both objects the temperatures (calculated from the luminosities of the stars) and the densities (derived from the mass-loss rates) are conducive for maser excitation.

For W Hya \citet{Justtanont_etal2005} modeled CO data to derive a temperature, $T$, profile, i.e., $T$ vs. distance from the star, $r$. For the maser region ($r = 1.8~10^{14}$ cm), one finds $T \approx 1000$ K from their Fig. 3. Also, assuming, like these authors, a luminosity of 5400 \Lsun\ and an effective temperature of 2500 K, respectively, one calculates a comparable 970 K at this $r$ from the Stefan-Boltzmann law.

The H$_2$ density  in a spherically symmetric outflow at distance $r$ from the central star is given by
$$n({\rm H}_2) = 1.13~10^{43}\Mdot(\Mspy)r^{-2}({\rm cm}){\rm v}_{\rm exp}^{-1}(\kms)$$
where \Mdot\ and ${\rm v}_{\rm exp}$ are the mass-loss rate and the expansion velocity, respectively, \citep[see, e.g. ][]{Millar1988,MentenAlcolea1995}.
If we assume for the masing region $n({\rm H}_2) = 10^9$ \ccm\ (see \S\ref{comparison}), $r = 1.6~10^{14}$ cm and an expansion velocity of 2 \kms\ and invert this equation to calculate the mass-loss rate, we obtain $\Mdot = 6~10^{-6}$ \Mspy, which is half the rate derived by \citet{ZubkoElitzur2000} and in between the estimates of \citet{Neufeld_etal1996} and \cite{Justtanont_etal2005} and, thus, hopefully, a plausible value.  \footnote{\citet{Neufeld_etal1996} derive a large value of $0.3$--$2~10^{-5}$ \Mspy\ for \Mdot\ from modeling
emission of thermally excited \hzo\ lines observed by the Infrared Space Observatory (ISO). In contrast, \citet{Justtanont_etal2005}, also using data from ISO as well as the Odin satellite, derive a much smaller
mass loss rate of  $(2.5\pm0.5)~10^{-7}$ \Mspy, comparable to values derived from CO lines. Their low \Mdot\ comes ``at the expense'' of an extremely high \hzo\ abundance of [\hzo/H$_2$ $=2~10^{3}$, which is significantly higher than the cosmic abundance of O. To explain it they invoke an influx of water from evaporating icy bodies in an Oort cloud analog, a mechanism that had earlier been advocated to explain the presence of water in IRC10216's envelope
\citep{Melnick_etal2001}. Although the calculations of Justtanont et al. make a strong case in favor of a low \Mdot\ one wonders whether a higher \Mdot\ and a lower \hzo\ abundance  in the range of the Neufeld et al. values might also be consistent with the measurements.}

Doing the same exercise for the RSG VY CMa with a luminosity of  $\approx 2~10^5$ \Lsun, we calculate for a region with an average radius of 300 AU (see \S\ref{spectralappearance}) a temperature of 480 K, which would be adequate at least for 22.2 GHz maser production. However, if we calculate the mass-loss rate required for a spherically symmetric outflow to maintain a density of $10^9$ \ccm\ at 300 AU, we obtain a value of 0.045 \Mspy. This is $\approx 200$ times higher than the generally quoted (already extremely high) mass-loss rate for this object (see \S\ref{intro}). We conclude that, in contrast to W Hya, at least the 22.2 GHz \hzo\ masers far from VY CMa cannot arise from the general material in a spherically symmetric outflow, but rather from special regions (e.g., shock fronts) within such a flow or from a collimated flow.
This situation is reminiscent of \hzo\ masers in star-forming regions modeled by \citet{Elitzur_etal1989}. These authors invoke a temperature (400 K) similar to the value we derive for VY CMa.

At this temperature, the highest excitation lines' maser emissivities are small (NM91), restricting their emission to regions closer to the star. This is in fact consistent with the smaller velocity ranges covered by the 321 and the 437 GHz lines.

\section{\label{summary}Summary}
We have used the APEX telescope for a multi-transition study of the AGB stars R Leo and W Hya and the red supergiant VY CMa. We find a variety of relative line ratios between the stars. For VY CMa we observe significant line shape differences between individual transitions, indicating that the lines' dominant contributions arise from different emission regions. While for all but one of the maser lines model calculation indeed predict inversion in a hot ($T = 1000$ K), dense ($n > 10^9$ \ccm) medium, other predictions of these calculations are at variance with the observations. This is not surprising, as key assumptions of the calculations do most likely not apply. We detected a new, high excitation maser line near 475 GHz that is predicted to only arise from a hot environment as described above.
Studying the excitation requirements and the extent of maser emission for W Hya and VY CMa, we conclude that (at least the 22.2 GHz) masers in the former may be excited in the  regular circumstellar outflow, while a special excitation mechanism is required for the much more widespread emission around VY CMa, possibly shocks in a higher velocity outflow.

\acknowledgements{We would like to thank the station manager, Alex Kraus, and
staff of the Effelsberg 100m telescope for taking and processing the 22.2 GHz
spectra.}

\bibliographystyle{aa}
\bibliography{8349}

\begin{thebibliography}{50}
\expandafter\ifx\csname natexlab\endcsname\relax\def\natexlab#1{#1}\fi

\bibitem[{{Cernicharo} {et~al.}(1997){Cernicharo}, {Alcolea}, {Baudry}, \&
  {Gonzalez-Alfonso}}]{Cernicharo_etal1997}
{Cernicharo}, J., {Alcolea}, J., {Baudry}, A., \& {Gonzalez-Alfonso}, E. 1997,
  \aap, 319, 607

\bibitem[{{Cernicharo} {et~al.}(1990){Cernicharo}, {Thum}, {Hein}, {John},
  {Garcia}, \& {Mattioco}}]{Cernicharo1990}
{Cernicharo}, J., {Thum}, C., {Hein}, H., {et~al.} 1990, \aap, 231, L15

\bibitem[{{Chen} {et~al.}(2000){Chen}, {Pearson}, {Pickett}, {Matsuura}, \&
  {Blake}}]{Chen2000}
{Chen}, P., {Pearson}, J.~C., {Pickett}, H.~M., {Matsuura}, S., \& {Blake},
  G.~A. 2000, \apjs, 128, 371

\bibitem[{{Cooke} \& {Elitzur}(1985)}]{CookeElitzur1985}
{Cooke}, B. \& {Elitzur}, M. 1985, \apj, 295, 175

\bibitem[{{Danchi} {et~al.}(1994){Danchi}, {Bester}, {Degiacomi}, {Greenhill},
  \& {Townes}}]{Danchi1994}
{Danchi}, W.~C., {Bester}, M., {Degiacomi}, C.~G., {Greenhill}, L.~J., \&
  {Townes}, C.~H. 1994, \aj, 107, 1469

\bibitem[{{de Jong}(1973)}]{deJong1973}
{de Jong}, T. 1973, \aap, 26, 297

\bibitem[{{Deguchi}(1977)}]{Deguchi1977}
{Deguchi}, S. 1977, \pasj, 29, 669

\bibitem[{{Elitzur} {et~al.}(1989){Elitzur}, {Hollenbach}, \&
  {McKee}}]{Elitzur_etal1989}
{Elitzur}, M., {Hollenbach}, D.~J., \& {McKee}, C.~F. 1989, \apj, 346, 983

\bibitem[{{G{\"u}sten} {et~al.}(2006){G{\"u}sten}, {Nyman}, {Schilke},
  {Menten}, {Cesarsky}, \& {Booth}}]{Gusten_etal2006}
{G{\"u}sten}, R., {Nyman}, L.~{\AA}., {Schilke}, P., {et~al.} 2006, \aap, 454,
  L13

\bibitem[{{Harwit} \& {Bergin}(2002)}]{HarwitBergin2002}
{Harwit}, M. \& {Bergin}, E.~A. 2002, \apjl, 565, L105

\bibitem[{{Herpin} {et~al.}(1998){Herpin}, {Baudry}, {Alcolea}, \&
  {Cernicharo}}]{Herpin_etal1998}
{Herpin}, F., {Baudry}, A., {Alcolea}, J., \& {Cernicharo}, J. 1998, \aap, 334,
  1037

\bibitem[{{Heyminck} {et~al.}(2006){Heyminck}, {Kasemann}, {G{\"u}sten}, {de
  Lange}, \& {Graf}}]{Heyminck_etal2006}
{Heyminck}, S., {Kasemann}, C., {G{\"u}sten}, R., {de Lange}, G., \& {Graf},
  U.~U. 2006, \aap, 454, L21

\bibitem[{{Ho} {et~al.}(2004){Ho}, {Moran}, \& {Lo}}]{Ho_etal2004}
{Ho}, P.~T.~P., {Moran}, J.~M., \& {Lo}, K.~Y. 2004, \apjl, 616, L1

\bibitem[{{Hunter} {et~al.}(2007){Hunter}, {Young}, {Christensen}, \&
  {Gurwell}}]{Hunter_etal2007}
{Hunter}, T.~R., {Young}, K.~H., {Christensen}, R.~D., \& {Gurwell}, M.~A.
  2007, ArXiv e-prints, 704

\bibitem[{{Josselin} \& {Plez}(2007)}]{JosselinPlez2007}
{Josselin}, E. \& {Plez}, B. 2007, \aap, 469, 671

\bibitem[{{Justtanont} {et~al.}(2005){Justtanont}, {Bergman}, {Larsson},
  {Olofsson}, {Sch{\"o}ier}, {Frisk}, {Hasegawa}, {Hjalmarson}, {Kwok},
  {Olberg}, {Sandqvist}, {Volk}, \& {Elitzur}}]{Justtanont_etal2005}
{Justtanont}, K., {Bergman}, P., {Larsson}, B., {et~al.} 2005, \aap, 439, 627

\bibitem[{{Klein} {et~al.}(2006){Klein}, {Philipp}, {Kr{\"a}mer}, {Kasemann},
  {G{\"u}sten}, \& {Menten}}]{Klein_etal2006}
{Klein}, B., {Philipp}, S.~D., {Kr{\"a}mer}, I., {et~al.} 2006, \aap, 454, L29

\bibitem[{{Knapp} {et~al.}(2003){Knapp}, {Pourbaix}, {Platais}, \&
  {Jorissen}}]{Knapp_etal2003}
{Knapp}, G.~R., {Pourbaix}, D., {Platais}, I., \& {Jorissen}, A. 2003, \aap,
  403, 993

\bibitem[{{Lada} \& {Reid}(1978)}]{LadaReid1978}
{Lada}, C.~J. \& {Reid}, M.~J. 1978, \apj, 219, 95

\bibitem[{{Liljestr{\"o}m} {et~al.}(2002){Liljestr{\"o}m}, {Winnberg}, \&
  {Booth}}]{Liljestrom_etal2002}
{Liljestr{\"o}m}, T., {Winnberg}, A., \& {Booth}, R. 2002, in IAU Symposium,
  Vol. 206, Cosmic Masers: From Proto-Stars to Black Holes, ed. V.~{Migenes} \&
  M.~J. {Reid}, 314--+

\bibitem[{{Melnick} {et~al.}(1993){Melnick}, {Menten}, {Phillips}, \&
  {Hunter}}]{Melnick_etal1993}
{Melnick}, G.~J., {Menten}, K.~M., {Phillips}, T.~G., \& {Hunter}, T. 1993,
  \apjl, 416, L37+

\bibitem[{{Melnick} {et~al.}(2001){Melnick}, {Neufeld}, {Ford}, {Hollenbach},
  \& {Ashby}}]{Melnick_etal2001}
{Melnick}, G.~J., {Neufeld}, D.~A., {Ford}, K.~E.~S., {Hollenbach}, D.~J., \&
  {Ashby}, M.~L.~N. 2001, \nat, 412, 160

\bibitem[{{Menten}(2000)}]{Menten2000}
{Menten}, K.~M. 2000, in From Extrasolar Planets to Cosmology: The VLT Opening
  Symposium ESO ASTROPHYSICS SYMPOSIA. Springer-Verlag, 2000, p. 78, ed.
  J.~{Bergeron} \& A.~{Renzini}, 78--+

\bibitem[{{Menten} \& {Alcolea}(1995)}]{MentenAlcolea1995}
{Menten}, K.~M. \& {Alcolea}, J. 1995, \apj, 448, 416

\bibitem[{{Menten} \& {Melnick}(1989)}]{MentenMelnick1989}
{Menten}, K.~M. \& {Melnick}, G.~J. 1989, \apjl, 341, L91

\bibitem[{{Menten} \& {Melnick}(1991)}]{MentenMelnick1991}
{Menten}, K.~M. \& {Melnick}, G.~J. 1991, \apj, 377, 647

\bibitem[{{Menten} {et~al.}(1990{\natexlab{a}}){Menten}, {Melnick}, \&
  {Phillips}}]{Menten_etal1990a}
{Menten}, K.~M., {Melnick}, G.~J., \& {Phillips}, T.~G. 1990{\natexlab{a}},
  \apjl, 350, L41

\bibitem[{{Menten} {et~al.}(1990{\natexlab{b}}){Menten}, {Melnick}, {Phillips},
  \& {Neufeld}}]{Menten_etal1990b}
{Menten}, K.~M., {Melnick}, G.~J., {Phillips}, T.~G., \& {Neufeld}, D.~A.
  1990{\natexlab{b}}, \apjl, 363, L27

\bibitem[{{Menten} {et~al.}(2006){Menten}, {Philipp}, {G{\"u}sten}, {Alcolea},
  {Polehampton}, \& {Br{\"u}nken}}]{Menten_etal2006}
{Menten}, K.~M., {Philipp}, S.~D., {G{\"u}sten}, R., {et~al.} 2006, \aap, 454,
  L107

\bibitem[{{Menten} \& {Young}(1995)}]{MentenYoung1995}
{Menten}, K.~M. \& {Young}, K. 1995, \apjl, 450, L67+

\bibitem[{{Millar}(1988)}]{Millar1988}
{Millar}, T.~J. 1988, in Astrophysics and Space Science Library, Vol. 146,
  Astrophysics and Space Science Library, ed. T.~J. {Millar} \& D.~A.
  {Williams}, 287--+

\bibitem[{{Muller} {et~al.}(2007){Muller}, {Dinh-V-Trung}, {Lim}, {Hirano},
  {Muthu}, \& {Kwok}}]{Muller_etal2007}
{Muller}, S., {Dinh-V-Trung}, {Lim}, J., {et~al.} 2007, \apj, 656, 1109

\bibitem[{{Neufeld} {et~al.}(1996){Neufeld}, {Chen}, {Melnick}, {de Graauw},
  {Feuchtgruber}, {Haser}, {Lutz}, \& {Harwit}}]{Neufeld_etal1996}
{Neufeld}, D.~A., {Chen}, W., {Melnick}, G.~J., {et~al.} 1996, \aap, 315, L237

\bibitem[{{Neufeld} \& {Melnick}(1990)}]{NeufeldMelnick1990}
{Neufeld}, D.~A. \& {Melnick}, G.~J. 1990, \apjl, 352, L9

\bibitem[{{Neufeld} \& {Melnick}(1991)}]{NeufeldMelnick1991}
{Neufeld}, D.~A. \& {Melnick}, G.~J. 1991, \apj, 368, 215

\bibitem[{{Ohnaka}(2004)}]{Ohnaka2004}
{Ohnaka}, K. 2004, \aap, 424, 1011

\bibitem[{{Palma} {et~al.}(1988){Palma}, {Green}, {Defrees}, \&
  {McLean}}]{Palma_etal1988}
{Palma}, A., {Green}, S., {Defrees}, D.~J., \& {McLean}, A.~D. 1988, \apjs, 68,
  287

\bibitem[{{Reid} \& {Menten}(1990)}]{ReidMenten1990}
{Reid}, M.~J. \& {Menten}, K.~M. 1990, \apjl, 360, L51

\bibitem[{{Richards} {et~al.}(1998){Richards}, {Yates}, \&
  {Cohen}}]{RichardsCohen1998}
{Richards}, A.~M.~S., {Yates}, J.~A., \& {Cohen}, R.~J. 1998, \mnras, 299, 319

\bibitem[{{Risacher} {et~al.}(2006){Risacher}, {Vassilev}, {Monje}, {Lapkin},
  {Belitsky}, {Pavolotsky}, {Pantaleev}, {Bergman}, {Ferm}, {Sundin},
  {Svensson}, {Fredrixon}, {Meledin}, {Gunnarsson}, {Hagstr{\"o}m},
  {Johansson}, {Olberg}, {Booth}, {Olofsson}, \& {Nyman}}]{Risacher_etal2006}
{Risacher}, C., {Vassilev}, V., {Monje}, R., {et~al.} 2006, \aap, 454, L17

\bibitem[{{Rudnitskij}(1987)}]{Rudnitskij1987}
{Rudnitskij}, G.~M. 1987, in IAU Symposium, Vol. 122, Circumstellar Matter, ed.
  I.~{Appenzeller} \& C.~{Jordan}, 267--268

\bibitem[{{Schwarzschild}(1975)}]{Schwarzschild1975}
{Schwarzschild}, M. 1975, \apj, 195, 137

\bibitem[{{Sopka} {et~al.}(1985){Sopka}, {Hildebrand}, {Jaffe}, {Gatley},
  {Roellig}, {Werner}, {Jura}, \& {Zuckerman}}]{Sopka1985}
{Sopka}, R.~J., {Hildebrand}, R., {Jaffe}, D.~T., {et~al.} 1985, \apj, 294, 242

\bibitem[{{Tsuji}(1964)}]{Tsuji1964}
{Tsuji}, T. 1964, Annals of the Tokyo Astronomical Observatory, 9

\bibitem[{{Valdettaro} {et~al.}(2001){Valdettaro}, {Palla}, {Brand},
  {Cesaroni}, {Comoretto}, {Di Franco}, {Felli}, {Natale}, {Palagi}, {Panella},
  \& {Tofani}}]{Valdettaro_etal2001}
{Valdettaro}, R., {Palla}, F., {Brand}, J., {et~al.} 2001, \aap, 368, 845

\bibitem[{{Wilson} {et~al.}(2005){Wilson}, {Beasley}, \&
  {Wootten}}]{Wilson_etal2005}
{Wilson}, T.~L., {Beasley}, A.~J., \& {Wootten}, H.~A. 2005, in Astronomical
  Society of the Pacific Conference Series, Vol. 344, The Cool Universe:
  Observing Cosmic Dawn, ed. C.~{Lidman} \& D.~{Alloin}, 232--+

\bibitem[{{Yates} \& {Cohen}(1996)}]{YatesCohen1996}
{Yates}, J.~A. \& {Cohen}, R.~J. 1996, \mnras, 278, 655

\bibitem[{{Yates} {et~al.}(1995){Yates}, {Cohen}, \& {Hills}}]{Yates_etal1995}
{Yates}, J.~A., {Cohen}, R.~J., \& {Hills}, R.~E. 1995, \mnras, 273, 529

\bibitem[{{Young}(1995)}]{Young1995}
{Young}, K. 1995, \apj, 445, 872

\bibitem[{{Zubko} \& {Elitzur}(2000)}]{ZubkoElitzur2000}
{Zubko}, V. \& {Elitzur}, M. 2000, \apjl, 544, L137

\end{thebibliography}


\end{document}